\documentclass[12pt]{article}
\usepackage{graphicx}
\newcommand{\beq}{\begin{equation}}
\newcommand{\eeq}{\end{equation}}
\newcommand{\bea}{\begin{eqnarray}}
\newcommand{\eea}{\end{eqnarray}}

\renewcommand{\b}{\beta}
\renewcommand{\a}{\alpha}

\def\fun#1#2{\lower3.6pt\vbox{\baselineskip0pt\lineskip.9pt
  \ialign{$\mathsurround=0pt#1\hfil##\hfil$\crcr#2\crcr\sim\crcr}}}
\def\sq{{\vbox {\hrule height 0.6pt\hbox{\vrule width 0.6pt\hskip 3pt
   \vbox{\vskip 6pt}\hskip 3pt \vrule width 0.6pt}\hrule height 0.6pt}}}
\begin{document}
\begin{titlepage}
\begin{flushleft}
       \hfill                      {\tt hep-th/0303068}\\
       \hfill                       FIT HE - 03-01 \\
       \hfill                       Kagoshima HE-03-1 \\
\end{flushleft}
\vspace*{3mm}
\begin{center}
{\bf\LARGE Newton's law in de Sitter brane \\ }
\vspace*{5mm}

\bigskip

{\large Kazuo Ghoroku\footnote{\tt gouroku@dontaku.fit.ac.jp}\\ }
\vspace*{1mm}
{
\large 
Fukuoka Institute of Technology, Wajiro, Higashi-ku}\\
{
\large 
Fukuoka 811-0295, Japan\\}
\vspace*{3mm}

{\large Akihiro Nakamura\footnote{\tt nakamura@sci.kagoshima-u.ac.jp}}\\
\vspace*{1mm}
{\large Department of Physics, Kagoshima University, Korimoto 1-21-35}\\
{\large Kagoshima 890-0065, Japan\\}
\vspace*{3mm}

{\large Masanobu Yahiro\footnote{\tt yahiro@sci.u-ryukyu.ac.jp} \\}
\vspace{1mm}
{
\large 
Department of Physics and Earth Sciences, University of the Ryukyus,
Nishihara-chou, Okinawa 903-0213, Japan \\}

\vspace*{10mm}

\end{center}

\begin{abstract}
Newton potential has been evaluated for the case of dS brane embedded 
in Minkowski, dS$_5$ and AdS$_5$ bulks. We point out that
only the AdS$_5$ bulk
might be consistent with the Newton's law from the brane-world
viewpoint when we respect 
a small cosmological constant
observed at present universe.
\end{abstract}
\end{titlepage}

\section{Introduction}

It is quite expectable to consider our four dimensional world
as a brane like the one proposed in \cite{RS1,RS2}.
The recent interest is the de Sitter (dS) brane due to the
observation 
of a small but finite cosmological constant in the present universe.
The 4d Newton's law is guaranteed also on dS brane by the confirmation of
the localization of graviton on this brane for wide range of bulk 
configurations, for example AdS$_5$ and dS$_5$ \cite{bre,BG} as in \cite{RS2}. 
The non-localized modes, which
are called as Kaluza-Klein (KK) modes, give corrections to the Newton 
potential. They are dependent on the configuration of the bulk space.
For the Randall-Sundrum (RS) brane, massive KK modes yields 
correction like $1/r^3$ \cite{RS2,Gidd,DL}, and it is nicely understood 
from an idea of AdS/CFT correspondence \cite{DL}. After that, corrections
to the Newton's law on the brane in the other bulk have been studied
\cite{ADDK,CE,IT,KT,NO,KTT}. However some points have not yet been made clear.

It is then of interest and importance to make more analysis about
what kind of corrections appear. 
Our purpose is to see the corrections coming from KK modes to Newton's law
in the case of dS brane for 5d Minkowski, AdS$_5$ and dS$_5$ bulks. 
In Section 2, we give our model and the 5d graviton propagator to
study the 4d Newton potential.
In Section 3, the gravitational potentials are examined under a reasonable
setting. Summary is given in the final section.

\section{Graviton propagator}

The five-dimensional gravitational action is obtained, in the
Einstein frame, as\footnote{
Definitions taken here are, $R_{\nu\lambda\sigma}^{\mu}
=\partial_{\lambda}\Gamma_{\nu\sigma}^{\mu}-\cdots$, 
$R_{\nu\sigma}=R_{\nu\mu\sigma}^{\mu}$ and $\eta_{AB}=$diag$(-1,1,1,1,1)$. 
Five dimensional suffices are denoted by capital Latin and 
four dimensional ones by Greek letters. 
}
\beq
    S = {1\over 2\kappa^2}\Bigg\{
      \int d^5X\sqrt{-G} (R -  2\Lambda + L_{m})
          +2\int d^4x\sqrt{-g}K\Bigg\}
        -{\tau}\int d^4x\sqrt{-g} , 
\label{action}
\eeq
where $1/2\kappa^2=M^3$ and $K$ denotes 
the extrinsic curvature on the boundary. Five and four dimensional metrics
are denoted as $G_{MN}$ and $g_{\mu\nu}$.
The Lagrangian density $L_{m}$ represents a contribution from matter, and 
not needed to construct a background metric. 
The last term shows a brane action. 
The Einstein equation derived from $S$ is solved under an assumption, 
\beq
 ds^2= A^2(y)\left\{-dt^2+a_0^2(t)\gamma_{ij}(x^i)dx^{i}dx^{j}\right\}
           +dy^2  \, \label{metrica},
\eeq
where coordinates parallel and transverse to a brane are 
denoted by $x^{\mu}=(t,x^i)$ and $y$, respectively. 
The brane is located at $y=0$. 
We restrict our interest here to the case of a
Friedmann-Robertson-Walker type
(FRW) universe. Then, the three-dimensional metric $\gamma_{ij}$
is described in Cartesian coordinates as
$
  \gamma_{ij}=(1+k\delta_{mn}x^mx^n/4)^{-2}\delta_{ij},  \label{3metric}  
$
where the parameter values $k=0, 1, -1$ correspond to 
flat, closed, and open universe, respectively.
The scale factors, $a_0(t)$ and $A(y)$, are obtainable from the Einstein 
equation \cite{bre}. 

A perturbed metric $h_{ij}$, representing graviton, is assumed to have a form 
\beq
 ds^2= A^2(y)\left\{
  -dt^2+a_0(t)^2[\delta_{ij}+h_{ij}(t,x^i,y)]dx^{i}dx^{j}
\right\}
           +dy^2  \, , 
\label{metricape}
\eeq
where the case $k=0$ is taken. In this case,
$a(t)=e^{\sqrt{\lambda}t}$ and $\lambda$ is
the 4d cosmological constant, which is written by other parameters 
as \cite{bre}
\beq
   \lambda = \kappa^4\tau^2/36+\Lambda/6 . \label{4cos}
\eeq 
A traceless and transverse component, $h$, of the perturbation is relevant to 
Newton's law on our brane and its corrections. 
Projecting the component out with conditions $h_i^i=0$ and
$\nabla_i h^{ij}=0$, its 5d propagator, $\Delta_{5}$, should
satisfy the following equation,
\beq
{\sq}_5 \Delta_{5} (x,y;x^{\prime},y^{\prime}) = {\delta^{4}(x - x^{\prime})
\delta(y - y^{\prime}) \over {\sqrt {-G}}}\ \,, 
\label{green}
\eeq
where 
\bea
 \sq_5\equiv {1\over \sqrt{-G}}\partial_N \sqrt{-G} G^{NL}\partial_L
       = {1\over A^2(y)}\sq_4 + 
        (\partial_y^2+{4\over A}(\partial_yA)\partial_y) \, \label{mlapl} 
\eea
and $\sq_4=-\partial_t^2 -3\dot{a_0}/a_0\partial_t+\partial_i^2/a_0^2$.

\vspace{.3cm}
When a new coordinate $z$ and a redefined propagator $\Delta(x,z;x',z')$ 
are introduced as $\partial z/\partial y=\pm A^{-1}$
and $\Delta_5=A(z)^{-3/2}\Delta A(z')^{-3/2}$, 
$\Delta$ is solved as
\beq
 \Delta(x,z;x^{\prime},z^{\prime})=u(0,z) \Delta_0(x,x') u(0,z')
+ \int_{m_0^2}^{\infty} {d m^2}\: u(m,z) \Delta_m(x,x') u(m,z')
\, ,
\label{expand-green}
\eeq
\beq
[-\partial_z^2+V(z)]u(m,z)=m^2 u(m,z)  \; ,
\label{Schroedinger}
\eeq
\bea
(\sq_4^2-m^2)\Delta_m
= {\delta^{4}(x - x^{\prime}) \over \sqrt{-g}} \,,
\label{coeff}
\eea
where $V(z)={9\over 4}(\partial_y A)^2+{3\over 2}A\partial_y^2 A$,
and $m$ corresponds the mass observed on the brane, 
as seen in Eq. (\ref{coeff}). The explicit form of the
4d propagator, $\Delta_m$, on AdS brane is not expressed here since we don't
use it.
The eigenmodes, the solutions of (\ref{Schroedinger}), 
consist of a zero mode $u(0,z)$ 
and continuum KK modes $u(m,z)$ 
with $m^2>m_0^2$, for the given bulks, 
where $m_0^2=9\lambda/4$ \cite{bre}. The normalization of $u(0,z)$ is given by
demanding 
\beq
\int_{z_0}^{\infty}dz~u(0,z)^2=1.
\eeq
This integration is easily performed numerically.
As for the KK mode $u(m,z)$, its normalization is obtained by imposing the 
following condition
\beq
\int_{z_0}^{\infty}dz~u(m,z)u(m',z)=\delta(m^2-m'^2).
\label{norm-cond}
\eeq
The explicit form of $u(m,z)$ can be obtained in terms of the two independent
and complex-conjugate solutions (denoted by $F_1$ and $F_2$ below)
of the equation 
(\ref{Schroedinger}), together with the boundary condition on a brane 
\cite{bre}, 
\beq
 u'(z_0)=-{\kappa^2\tau\over 4}u(z_0).   \label{boundzero}
\eeq
The result is summarized as 
\beq
u(m,z)= { 1  \over 2i}  \sqrt{{1 \over \pi \a }} 
[e^{i\delta_0(\a)} F_1(z)-e^{-i\delta_0(\a)} F_2(z)] \; ,
\label{KK-wf}
\eeq
\beq
e^{2i\delta_0(\a)}={ {F_2'(z_0) + {\kappa^2 \tau \over 4} F_2(z_0)} 
           \over 
         {F_1'(z_0) + {\kappa^2 \tau \over 4} F_1(z_0)} } \,, 
\label{delta}
\eeq 
where $'=\partial/\partial_z$ and 
\bea
F_1(z)=Y^{-id} {}_2F_1(b_1,b_2;b_3;-Y), \quad
F_2(z)=Y^{id} {}_2F_1(b_1',b_2';b_3';-Y) \quad
{\rm for} \;\; \Lambda<0 \;, 
\label{sol-AdS} 
\\
F_1(z)=X^{-id} {}_2F_1(b_1,b_2;b_3;X),  \;\;\quad
F_2(z)=X^{id} {}_2F_1(b_1',b_2';b_3';X)  \;\;\;\quad
{\rm for} \;\; \Lambda>0 \;, 
\label{sol-dS}
\eea
\beq  
 Y={1\over\rm{sinh}^2(\sqrt{\lambda}z)}, \quad 
 X={1\over\rm{cosh}^2(\sqrt{\lambda}z)}, \quad 
 d={\sqrt{-9+4m^2/\lambda}\over 4},  
 \label{para1}
\eeq
\beq
  b_1=- {3\over 4}-id, \quad b_2={5\over 4}-id, \quad b_3=1-2id, 
\label{para2}
\eeq
\beq
  b_1'=- {3\over 4}+id, \quad b_2'={5\over 4}+id, \quad b_3'=1+2id . 
\label{para3}
\eeq
Here ${}_2F_1(b_1,b_2;b_3;X)$ denotes the Gauss's hypergeometric function. 
For $m>m_0$, $F_1(z)$ represents an outgoing wave asymptotically, 
while $F_2(z)$ does an incoming wave, and both are complex 
conjugate to each other. 
To see that this solution satisfies the above normalization condition,
it is convenient to use the followings two relations. The first one
is the following asymptotic form at large $z$, 
\beq
   u(m,z) \to \sqrt{{1 \over \pi \a }} \sin{(\a z + \delta_\a)} ,
\label{asymptotic-bc}
\eeq
where $\a=\sqrt{m^2-m_0^2}$, $\delta_\a$ means a phase dependent on $\a$.
The second relation is given by using (\ref{Schroedinger}) as, 
\beq
u(m,z)u(m',z)={1\over m^2-m'^2}
     \left\{u(m,z)\partial_z^2u(m',z)-u(m',z)\partial_z^2u(m,z)\right\}.
\eeq

The present universe implies 
a small $\lambda$, then 
we concentrate our discussion on such a case.
An observational time $t$ of Newton's law  
is much smaller than the cosmic age, $1/\sqrt{\lambda} \sim 10$ Gyr. 
For the case of such a small time, the scale factor 
$a_0=\exp{\sqrt{\lambda}t}$ on a brane 
is well approximated by $a_0=1$, and the de-Sitter propagator denoted above
by $\Delta_m$ can be approximated into the one in Minkowski space. 
The approximate form of the 5d propagator, 
valid at $\sqrt{\lambda}|t-t'| \ll 1$, is then obtained as 
\bea
 \Delta(x,z;x^{\prime},z^{\prime}) &=&
u(0,z)u(0,z')
\int {d^4p \over (2\pi)^4} {e^{ip(x-x')} \over -p^2 + i \epsilon } \nonumber \\
&& + \int_{m_0^2}^{\infty} {d m^2} \: u(m,z) u(m,z') 
\int {d^4p \over (2\pi)^4} {e^{ip(x-x')} \over -p^2-m^2 + i \epsilon } . 
\label{approx-expansion}
\eea
In the limit of $\lambda=0$, Eq. (\ref{approx-expansion}) is correct.

\section{Corrections to Newton's Law}

The static potential $\tilde U(r)$ between two objects of unit mass on a brane is defined as \cite{Gidd}
\bea
    U(r) &=& \tilde U(r)/\kappa^2 = - \int_{-\infty}^{\infty} dt 
           \Delta_{5}(t,x_i,y;t',x_i',y')|_{y=y'=0, t'=0} \nonumber \\
         &=& - \int_{-\infty}^{\infty} dt 
           \Delta(t,x_i,z;t',x_i',z')|_{z=z'=z_0, t'=0} , 
\label{static-pot}
\eea
where $r=|\vec{x}-\vec{x'}|$. 
Inserting the approximate form (\ref{approx-expansion}) into 
Eq. (\ref{static-pot}) leads to $U(r) = U_0 + \Delta U$ for 
\beq
U_0 \equiv { u(0,z_0)^2 \over 4 \pi r}
\;, \quad 
 \Delta U \equiv \int_{m_0^2}^{\infty} {d m^2} \: 
 u(m,z_0)^2 {e^{-mr} \over 4 \pi r} \; .
\label{static-approx1}
\eeq
The term $U_0$ guarantees Newton's law, and $\Delta U$ represents 
its correction. The correction depends 
on the magnitude of KK mode $u(m,z_0)$ on a brane.

Particularly at $z=z_0$, namely on a brane, the KK mode has a simple form 
\beq
u(m,z_0)= - \sqrt{{1 \over \pi \a }} 
{ \a \over |F_1'(z_0) + {\kappa^2 \tau \over 4} F_1(z_0)| }, 
\label{KK-wf0}
\eeq
where use has been made of 
$F_1'(z)F_2(z)-F_2'(z)F_1(z)=2i\a$. Inserting 
Eq. (\ref{KK-wf0}) into Eq. (\ref{static-approx1}) leads to 
\beq
 \Delta U = {1 \over 2 \pi^2 \: r } \int_{m_0}^{\infty} {d m} \: 
 { m \a \over F } e^{-mr}  \; , 
\quad F \equiv |F_1'(z_0) + {\kappa^2 \tau \over 4} F_1(z_0)|^2.
\label{static-approx2}
\eeq
Equation (\ref{static-approx2}) thus obtained is based on 
Eq. (\ref{approx-expansion}) which is valid 
for $\sqrt{\lambda}|t-t'| \ll 1$. So Eq. (\ref{static-approx2}) is accurate 
for $\sqrt{\lambda}r \ll 1$, because the distance $r$ between two massive 
objects 
is related to the propagation time of graviton $|t-t'|$ as 
$r \approx |t-t'|$. Particularly in the limit $\lambda \to 0$, 
Eq. (\ref{static-approx2}) is correct for any $r$. 

The correction (\ref{static-approx2}) is different from the corresponding one 
in Ref. \cite{KT}, since the normalization and boundary conditions, 
(\ref{norm-cond}) and (\ref{boundzero}), are not imposed there  \cite{KT1}. 
Then their results could not reproduce the $1/r^3$ correction in the limit
of $\lambda \to 0$. In our case, it can be seen as shown below.

\subsection{Randall-Sundrum brane} 

As for the Randall-Sundrum brane, in which $\lambda=0$,   
corrections to Newton's law are well known at $r \gg L$ \cite{RS2}, 
where $L$ is the radius defined by $L=\sqrt{6/|\Lambda|}$. 
In this subsection, the corrections are analyzed for both regions 
of $r < L$ and $r > L$. 

\vspace{.3cm}
For $\lambda=0$, 
the corresponding solutions $F_1(z)$ and $F_2(z)$ of 
the equation (\ref{Schroedinger}) are given as 
$F_1(z)=\sqrt{\pi mz/2}H_2^{(1)}(m z)$ and 
$F_2(z)=\sqrt{\pi mz/2}H_2^{(2)}(m z)$, 
where $H_n^{(1,2)}(x)=J_n(x) \pm iN_n(x)$ for the Bessel functions 
$J_n(x)$ and $N_n(x)$ of integer $n$. Since $z_0=L$ in the case,
$F$ in Eq. (\ref{static-approx2}) is expressed as 
$F=\pi m^3 L | H_1^{(1)}(m L) |^2 /2 $, indicating that 
$F \to 2m/(\pi L)$ at the small limit of $mL$ and $F \to m^2$ at the large 
limit. 

Firstly, consider the region $r \gg L$ where $F$ is approximated
by the one for small $mL$ since $m$ and $r$ are 
mutually conjugate due
to the factor $e^{-mr}$ in the integrand in (\ref{static-approx2}). 
Then we obtain
$$\Delta U\sim L/(4 \pi r^3),$$
which leads to a well-known result $\Delta U/U_0=L^2/(2r^2)$ \cite{DL}. 

\vspace{.2cm}
While in the region of small $r$, $L\gg r$, the potential can be estimated
by the approximation of $F\sim m^2$, and we obtain
$$\Delta U \sim 1/(2 \pi^2 r^2).$$ 
It indicates $\Delta U/U_0=L/(\pi r) \gg 1$, then 
the pole contribution of $1/r$ is small and 
5d Newton's law appears as the dominant potential
in the region $ r \ll L$ as expected. This is pointed out also in \cite{ADDK,CE,IT,NO}.

\subsection{dS brane with small $\lambda$} 

The dS brane, in which $\lambda>0$, can be embedded in three types of 
bulks, AdS$_5$ \cite{K, BDEL}, dS$_5$ \cite{BDEL} and the 5d Minkowski 
space \cite{KL}. In this case, two scale parameters appear in studying 
the potential at some region of $r$. Due to the
relation (\ref{4cos}), the region of $r$ in the two bulks, dS$_5$ and 
the 5d Minkowski space, is restricted to a short range region as shown below.

\vspace{.2cm}
For the 5d Minkowski space as a simple case, $F_1$ is obtained as 
$F_1=\exp{(i \a z)}$, leading to $F=m^2$. This is understood also from
the above approximate form of $F$ at large $mL$. For 5d Minkowski, $L=\infty$
and this implies that the potential in this case expresses the exact 5d limit
at any $r$.

\vspace{.2cm}
The similar situation is seen also in
the dS bulk. Consider it with the radius $L=\sqrt{6/\Lambda}$. 
As noted above in (\ref{4cos}), $\lambda$ is related to $\Lambda$ as 
$\lambda = \kappa^4\tau^2/36+\Lambda/6$. This relation shows  
that $\lambda > \Lambda/6 $ for the dS bulk, that is, $1\leq m/m_0<mL$. 
Then we can not see the region of $L \ll r$ or $mL\ll 1$. Therefore,
$1/r^3$ correction can not be seen in this case. And the available region
is restricted to the short range region. As a result, 
it is easy to see 5d potential, $1/r^2$, at small $r$ as in the case of RS
given above.

Further, we can see that the potential $1/r$ coming from the 
trapped zero mode
is not the leading term any more in this region.
The $1/r$ term $U_0$ depends on the magnitude $u(0,z_0)$ 
on the brane. 
The magnitude is estimated by using the explicit form of the solutions given 
above (\ref{sol-dS}). By setting as $m=0$ we obtain, 
$u(0,z_0)=d_1 F_1(z_0)$ with $F_1(z)=(\cosh{{\sqrt{\lambda}z}})^{-3/2}$ 
and $d_1=\langle F_1|F_1\rangle_z^{-1/2}$. 
It is impossible to integrate $F_1^2$ over $z$ analytically. 
So an order estimate is made for $d_1$ with the relation, 
$\{\exp{({\sqrt{\lambda}z})}\}^{-3} < F_1^2< 
\{\exp{({\sqrt{\lambda}z})}/2\}^{-3}$, 
which is valid for any positive $z$. Making an analytic integration 
for each function in the lower and upper bounds of the relation, 
one can obtain the relation 
$\sqrt{m_0/4} f^{3/2}<u(0,z_0)<\sqrt{2m_0}f^{3/2}$, 
where $f(\b)=1+\sqrt{1-\b^2}$ for 
$\b \equiv \sqrt{\Lambda/6\lambda}=3/(2Lm_0)$. 
This indicates that $u(0,z_0)$ is of order $\sqrt{m_0}$, because 
$1<f<2$ in the entire region $0<\b<1$. 
This order estimation leads to $U_0 \approx m_0/(4\pi r)$ 
for the case of dS bulk.

The terms $U_0$ and $\Delta U$ calculated above lead to 
$\Delta U/U_0 \approx 1/(rm_0) \gg 1$ for $r \ll 1/m_0$. 
From the observation or our assumption $\lambda\sim m_0^2 \ll M^2$, 
this estimation is justified. This indicates 
that the present universe is not embedded in the dS 
or Minkowski bulk when we consider according to our brane model.

\vspace{.5cm}
As for AdS$_5$, (\ref{4cos}) gives no considerable constraint on $mL$ or the
range of $r$. As a matter of fact, the situation is similar to the case 
of the RS brane since $m_0$ is so small compared to the value of $1/L$.
However $F$ is slightly
different from the one of RS, especially near $m=m_0$. While
the function $F$ has the same form in the large 
limit of $m^2/|\Lambda|$, since $F_1$ tends to $\exp{(i \a z)}$ in the limit. 
Figure 1 shows the behavior of $F$ at smaller $m$ for various $\lambda$. 
\begin{figure}[htbp]
\begin{center}
\voffset=15cm
  \includegraphics[width=8cm,height=7cm]{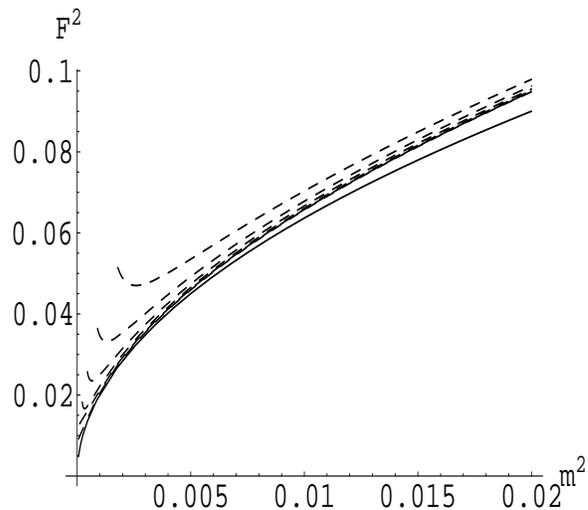}
\caption{The solid curve shows ${2\over \pi} m$, 
and the dotted curves represent
$F$ for $\mu=1$ and $\lambda=10^{-2.8-0.3N}$ where $N=1\sim 6$ from
the highest to the lowest one. The end points at small $m$ of the dotted curves
correspond to the values at $m=m_0$ for each $\lambda$.
We can see that $F$ approaches to
the RS limit as $\lambda\to 0$.}
\end{center}
\end{figure}
The value of $\lambda$ can be measured by $\Lambda$ in the theory, 
so we take as $10^{-3.1}<\lambda/|\Lambda|<10^{-4.6}$ in the figure. But
it should be taken at about $\sim 10^{-15}$ actually, then the realistic
case is infinitely near the RS limit. 

\vspace{.2cm}
Generally, it is possible to approximate $F$ by 
\beq
F \approx c_0 + c_1 m + c_2m^2 \; , 
\label{F-approx-dS} 
\eeq
where $c_i$ are dependent on $\mu=\sqrt{-\Lambda/6}$ and $\lambda$.
When $\lambda$ becomes small, $F$ approaches to the one of RS brane, since 
$c_0\to 0$, $c_1\to 2/\pi L$, $c_2\to 1$ in the limit. 
So it would be possible to find a
similar potential at large $r$ to the one of RS case when the parameters 
$\lambda$ and
$\Lambda$ are appropriately chosen. 

However the essential difference
from the RS case would be seen in the ratio $\Delta U/U_0$, which represents
the ratio of the correction like $1/r^3$ and the leading term of $1/r$.
It would be expected that this ratio shifts from the RS case, 
$\Delta U/U_0=L^2/(2 r^2)$, and can be written as
$$\Delta U/U_0=f(L,\lambda)/(2 r^2). $$
To study the meaning of this difference is an interesting
problem from the theoretical viewpoint of AdS/CFT correspondence. 
We will discuss this issue in the future paper.

\vspace{.2cm}
As a result of this section, we can say that the favorable bulk
configuration of the brane-world would be the AdS${}_5$

\vspace{.5cm}

\section{Summary}

In this paper, Newton potential has been evaluated for the case of dS 
brane embedded in Minkowski, dS$_5$ and AdS$_5$ bulks.  

For this purpose, an approximate propagator (\ref{approx-expansion}) 
has been derived, which is valid at $\sqrt{\lambda}|t-t'|\ll 1$.  
Then on the basis of the propagator, the static potential $U(r)$ has been  
divided into $U_0$ which guarantees Newton's law and $\Delta U$ which 
represents its correction.  The formula (\ref{static-approx2}), which 
is accurate for $\sqrt{\lambda}r\ll 1$, was used to evaluate the correction.  

Next it is verified to reproduce the correct RS limit in the case of 
$\lambda=0$ \cite{DL,IT,NO}.  
Then the case of 5d Minkowski was examined and it was shown that 
the potential expresses the exact 5d limit at any $r$.  
The similar situation was seen also in the dS bulk.  Namely we can not 
see the region of $L\ll r$ so that the available region is restricted to 
the short range region.  Furthermore, we could see the potential 
$U_0\sim 1/r$ is not leading term but the ``correction'' 
$\Delta U\sim 1/r^2$ is the dominant part.  This indicates
that the present universe is not embedded in the Minkowski or dS bulk 
when we consider according to our brane model.  

As for AdS$_5$ bulk, it would be possible to find the similar potential 
at large $r$ to the one of RS case when the parameters $\lambda$ and
$\Lambda$ are appropriately chosen.
However the essential difference lies in the fact that it would be 
expected that the ratio $\Delta U/U_0$ shifts from the RS case,
$\Delta U/U_0=L^2/(2 r^2)$, and can be written as
$\Delta U/U_0=f(L,\lambda)/(2 r^2)$.  
To study the meaning of this difference is an interesting
problem from the theoretical viewpoint of AdS/CFT correspondence.
We will discuss this issue in the future paper.

In conclusion, we can say that the favorable bulk
configuration of the brane-world would be the AdS${}_5$ 
at the present universe.  The formula (\ref{static-approx2}) valid at 
$\sqrt{\lambda}r\ll 1$ is useful in comparing this theory 
with the measured corrections to Newton's law, 
because all the measurements are performed 
in the region $\sqrt{\lambda}r\ll 1$.


\section*{Acknowledgments}
This work has been supported in part by the Grants-in-Aid for
Scientific Research (13135223, 14540271)
of the Ministry of Education, Science, Sports, and Culture of Japan.


\begin{thebibliography}{99}
\bibitem{KT1} The solutions $\psi$ given by Eq.(25) in \cite{KT}
do not satisfy the 
boundary condition, $\psi ' (0_+)/\psi(0)=-\sigma/(12M^3)$, 
required from (24) in \cite{KT}. If the $\psi$ are inserted 
on the LHS of the condition, it changes with the mass $m$ of KK modes, while 
the RHS does not. Thus, the $\psi$ have a wrong $m$ depencence. 
The normalization for $\psi$ is also 
obscure in \cite{KT}. The normalization factor depends in general
on $m$. Then $\psi$ used in their analyses should change 
the $r$-dependence of the potential. These points would be essential
in deriving the correction for $V(r)$.
\end{thebibliography}
\end{document}